\documentclass[lettersize,journal]{IEEEtran}
\usepackage{amsmath,amsfonts}
\usepackage{algorithmic}
\usepackage{algorithm}
\usepackage{array}
\usepackage[caption=false,font=footnotesize]{subfig}
\usepackage{textcomp}
\usepackage{stfloats}
\usepackage{url}
\usepackage{verbatim}
\usepackage{graphicx}
\usepackage{booktabs}
\usepackage{cite}
\usepackage{times}
\usepackage{epsfig}
\usepackage{multirow}
\usepackage{multicol}
\usepackage[table]{xcolor}
\usepackage{yhmath}
\usepackage{bm}
\usepackage{pifont}
\usepackage{eso-pic}
\usepackage{bbding}
\usepackage[autostyle=true]{csquotes}

\usepackage[colorlinks,linkcolor=red]{hyperref}

\usepackage{tikz}

\usepackage{seqsplit} 
\usepackage{placeins} 

\definecolor{lime}{HTML}{A6CE39}
\DeclareRobustCommand{\orcidicon}{%
    \begin{tikzpicture}
    \draw[lime, fill=lime] (0,0) 
    circle [radius=0.16] 
    node[white] {{\fontfamily{qag}\selectfont \tiny ID}};    \draw[white, fill=white] (-0.0625,0.095) 
    circle [radius=0.007];    \end{tikzpicture}
    \hspace{-2mm}}
\foreach \x in {A, ..., Z}{%
    \expandafter\xdef\csname orcid\x\endcsname{\noexpand\href{https://orcid.org/\csname orcidauthor\x\endcsname}{\noexpand\orcidicon}}
    }

\definecolor{myred}{HTML}{ff9999}
\definecolor{mygreen}{HTML}{99ff99}
\definecolor{myblue}{HTML}{9999ff}
\definecolor{avgcolumnbg}{HTML}{F2F2F2}

\begin{document}

\title{KD-NVC: A Search-and-Distill Framework to Accelerate Neural Video Coding}

\markboth{This manuscript is prepared for submission to IEEE Transactions }
{This manuscript is prepared for submission to IEEE Transactions }

\IEEEpubid{0000--0000/00\$00.00~\copyright~2026 IEEE}
        
\author{
Yuxiao Sun\orcidA{}, 
Meiqin Liu\orcidC{}, 
Chao Yao\orcidD{},~\IEEEmembership{Member,~IEEE}, 
Hui Xiang\orcidI{},
Jingran Wu,
Xianguo Zhang\orcidH{}\\
Jian Jin\orcidE{},~\IEEEmembership{Member,~IEEE}, 
Weisi Lin\orcidG{},~\IEEEmembership{Fellow,~IEEE}, 
and Yao Zhao\orcidB{},~\IEEEmembership{Fellow,~IEEE}, 
\thanks{
This work was supported in part by the National Natural Science Foundation of China under Grant 62120106009, Grant 62372036, Grant U24B20179, and Grant 62332017.
\textit{(Corresponding author: Yao Zhao.)}

Yuxiao Sun, Meiqin Liu, and Yao Zhao are with the Institute of Information Science, Beijing Jiaotong University, Beijing 100044, China, and also with the Visual Intelligence +X International Cooperation Joint Laboratory of MOE, Beijing 100044, China (e-mail: \url{yuxiaosun@bjtu.edu.cn}; \url{mqliu@bjtu.edu.cn}; \url{yzhao@bjtu.edu.cn}).

Chao Yao is with the School of Computer and Communication Engineering, University of Science and Technology Beijing, Beijing 100083, China (e-mail: \url{yaochao@ustb.edu.cn}).

Hui Xiang, Jingran Wu, and Xianguo Zhang are with Tencent, Beijing 100193, China (e-mail: \url{ihuixiang@tencent.com}; \url{jingranwu@tencent.com}; \url{codeczhang@tencent.com}).

Jian Jin and Weisi Lin are with the School of Computer Science and Engineering, Nanyang Technological University, Singapore 639798 (e-mail:
\url{jian.jin@ntu.edu.sg}; \url{wslin@ntu.edu.sg}).
}
}

\maketitle

\begin{abstract}
While neural video coding (NVC) has achieved remarkable rate-distortion performance, real-time decoding on edge devices has become an important demand but remains limited by high complexity. Knowledge distillation (KD) is widely used for model acceleration, yet its application to NVC faces critical challenges.
Specifically, the heterogeneity of NVC sub-modules renders uniform architectural reduction suboptimal, necessitating a per-module design for better rate-distortion-speed trade-off. However, searching for diverse architectures via existing neural architecture search (NAS) algorithms is unaffordable due to the expensive training cost of neural video codecs. Moreover, after the lightweight architecture is determined, existing distillation methods overlook the feature-energy sparsity induced by the rate-constraint, which is essential for maintaining compression performance.
To address these issues, we propose a two-stage distillation framework KD-NVC. In the first stage, we introduce an acceleration-efficiency-based neural architecture search (AE-NAS) algorithm. It explores the module-wise Pareto frontier to adaptively allocate the acceleration budget across heterogeneous modules. Also, it introduces the acceleration-efficiency metric to determine the final student architecture without practically training all architecture-level candidates. In the second stage, we design an energy-aware feature distillation (EFD) loss that aligns the spatially-aggregated feature-energy signatures between the teacher and student codecs, transferring the rate-induced sparsity patterns for better compression efficiency. Experimental results demonstrate that the proposed framework consistently outperforms existing codec-oriented distillation methods, and achieves 69 FPS decoding at 1080p on RTX 5060 while maintaining comparable RD performance to VTM-LDB. 

\end{abstract}


\section{Introduction}

\begin{figure}
    \centering
    \includegraphics[width=\linewidth]{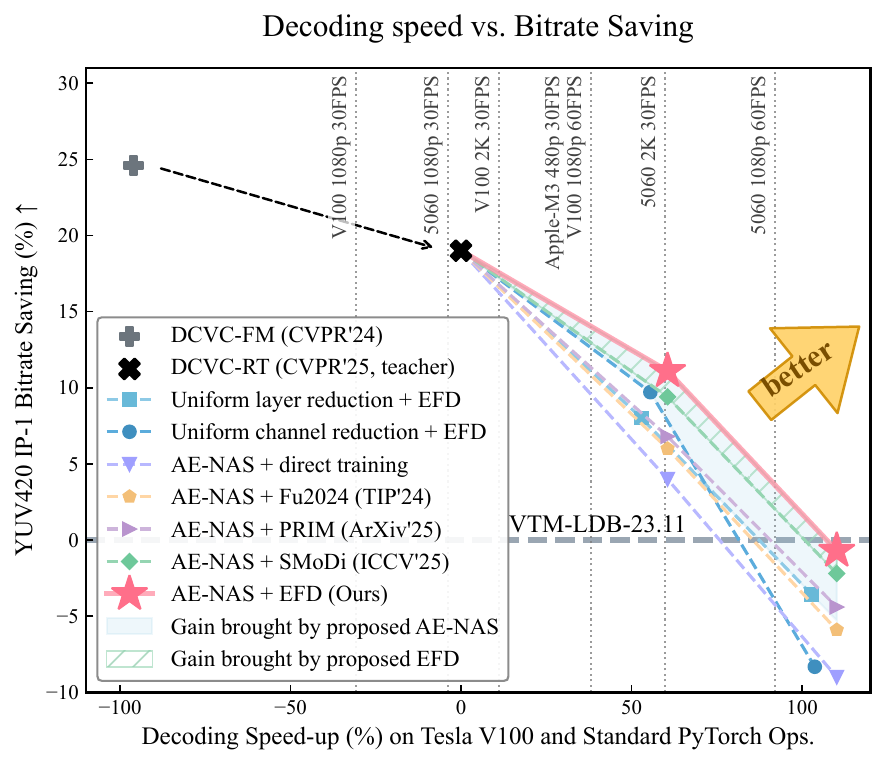}
    \caption{Acceleration and rate-distortion performance of compared methods. Both the proposed acceleration-efficiency-based neural architecture search (AE-NAS) and energy-aware feature distillation (EFD) loss achieve promising performance improvement.}
    \label{figure_first}
\end{figure}

\IEEEPARstart{N}{eural} video coding (NVC) has recently achieved remarkable progress with the variational auto-encoder (VAE)~\cite{autoencoder} architecture and end-to-end optimization. Many methods~\cite{dcvc_fm, dcvc_rt, nvc1b} achieve superior rate-distortion~(RD) performance compared with standard video codecs~\cite{h265, h264, vtm}. 
Nevertheless, practical deployment of existing methods on high-resolution (e.g., 2K), high-frame-rate (60 FPS), and resource-constrained scenarios remains challenging, as shown in Fig.~\ref{figure_first}.

\IEEEpubidadjcol
Due to the inherent characteristics of VAE structures, most existing neural video codecs exhibit symmetric complexity between the encoder and decoder~\cite{dvc, dcvc_dc, dcvc_fm, dcvc_rt, ecvc}. For diffusion-based codecs~\cite{i2vc, gnvc_vd}, the decoder complexity far exceeds that of the encoder. However, such symmetric or decoder-heavy designs mismatch the requirements of most practical applications~\cite{h265, asym_llic}. For example, video distribution typically follows a ``one-to-many'' paradigm, where encoding is a one-time process offloaded to high-performance servers, while decoding is executed repeatedly on many resource-limited devices. This mismatch drives an urgent need for decoder-side model acceleration.

\begin{figure}[!t]
    \centering
    \captionsetup[subfloat]{font=footnotesize,justification=justified,singlelinecheck=false,format=plain,indention=0pt}
    \subfloat[Neural video codecs have a longer and more complex inference chain, where the functions of different modules are more diverse.]{%
        \includegraphics[width=\linewidth]{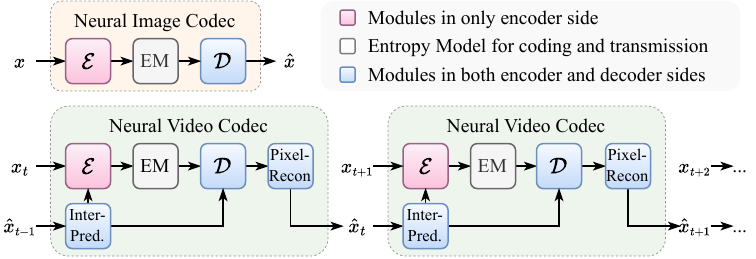}%
    }\\[2pt]
    \subfloat[Existing progressive and serial paradigms have low complexity but suffer from prohibitively long training time (over one month with adequate V100 GPUs), while exhaustive paradigms can be fully parallelized but require an unaffordable number of trained candidates (over $10,000$ GPU days if there are $100$ candidates); both are impractical for neural video codec training.]{%
        \includegraphics[width=\linewidth]{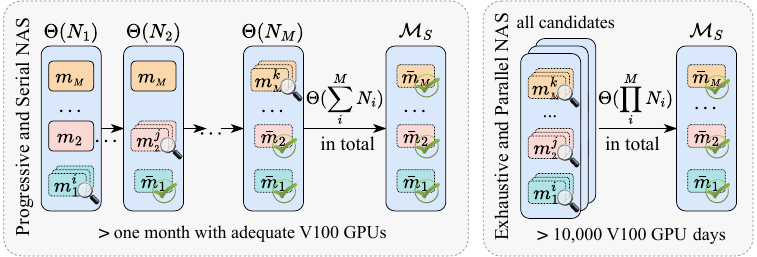}%
    }\\[2pt]
    \subfloat[Channel-wise energy distributions of intermediate features ($L_1$/$L_2$/$L_3$, from shallow to deep) in neural video codecs (DCVC-RT~\cite{dcvc_rt} and DVC~\cite{dvc}, w/ and w/o the rate constraint) and pretrained visual backbones~\cite{vgg, resnet, autoencoder}. Energy-sparsity is computed by the top-10\% energy ratio ($T10$)\protect\footnotemark{} and the normalized Gini coefficient ($G$). Compared with common visual backbones, neural video codecs exhibit a uniquely sparse and unbalanced channel-wise energy distribution that is induced by the bitrate constraint.]{%
        \includegraphics[width=\linewidth]{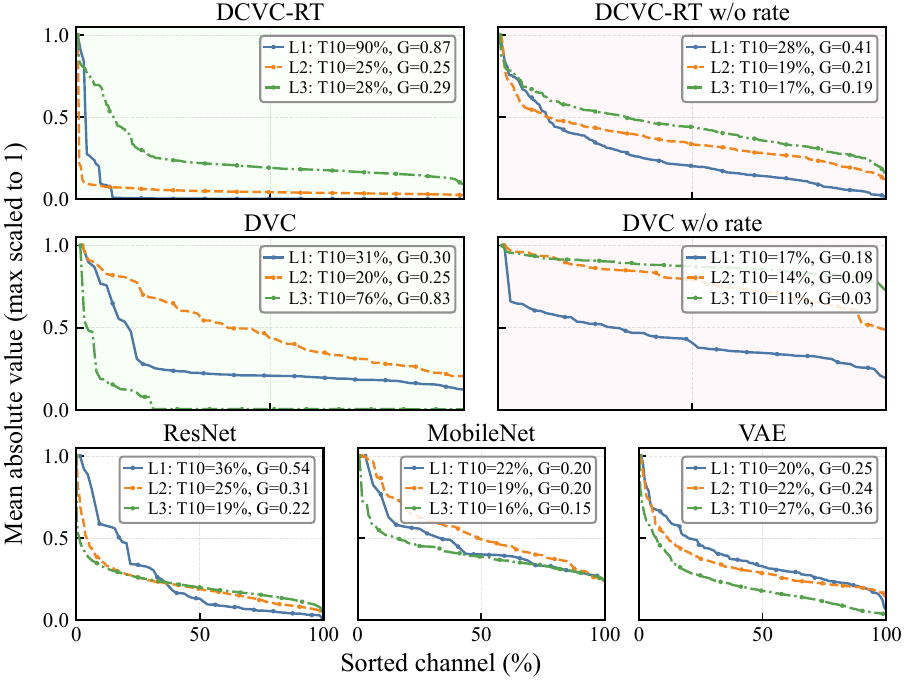}%
    }
    \caption{Three key observations about using NAS and KD on video coding.}
    \label{figure_observation}
\end{figure}

Knowledge distillation (KD) is a widely used technique for model acceleration, which transfers knowledge from a large teacher model to a compact student model. In learned image compression, KD has been mainly used to accelerate the codec~\cite{fu2024tip, prim, iccv2025smodi}. Meanwhile, some methods~\cite{unicompress, free_vsc, smc++, sec_vcm, s2vc} use KD to transfer semantic-level knowledge from pretrained visual foundation models to neural codecs. 
However, KD for speeding up neural video coding remains underexplored and faces three primary challenges:

\label{section_observation}

\textit{\textbf{Observation \MakeUppercase{\romannumeral 1}: Module heterogeneity encourages per-module lightweight design.}} Compared to image coding, video coding has a more complex inference chain within a single frame and a longer temporal reference chain over multiple frames, as shown in Fig.~\ref{figure_observation}~(a). Different modules in the video codec serve distinct roles (e.g., inter-prediction, contextual coding, entropy coding), which leads to different sensitivities to model compression.
Such heterogeneity implies that handcrafted uniform layer or channel reduction overlooks module-wise differences, leading to suboptimal rate-distortion-speed trade-offs, as shown in Fig.~\ref{figure_first}.
Thus, leveraging neural architecture search (NAS), which searches over a predefined candidate space to determine final student architectures, is essential for finding a per-module design.

\footnotetext{For an intermediate feature $F\in\mathbb{R}^{C\times H\times W}$, we define feature energy as the non-negative channel response magnitude, computed as $e_c=\frac{1}{HW}\sum_{h,w}|F_{c,h,w}|$. $T10$ is the fraction of $\sum_c e_c$ carried by the top-10\% highest-energy channels, and $G$ is the normalized Gini coefficient computed over $\{e_c\}_{c=1}^{C}$.}

\textit{\textbf{Observation \MakeUppercase{\romannumeral 2}: NAS cost needs to be reduced.}} 
Existing NAS paradigms cannot be directly employed for neural video codecs. Training SOTA video codecs is ``expensive'', typically taking millions of iterations and dozens of GPU days~\cite{dcvc_tcm, dcvc_sdd, dcvc_rt}, which makes existing progressive, multi-stage, and exhaustive NAS paradigms unaffordable for video codec training, as shown in Fig.~\ref{figure_observation}~(b). It creates an urgent need for a paradigm that can simplify the NAS process and efficiently determine the final architecture to avoid heavy time cost.

\textit{\textbf{Observation \MakeUppercase{\romannumeral 3}: Rate-constraint-induced feature-energy sparsity needs to be addressed.}} 
Unlike common visual backbones, the intermediate features of neural video codecs exhibit highly sparse and unbalanced channel-wise energy distributions~\cite{iccv2025smodi, minnen2018joint, soap_lic}, where a few channels carry large magnitudes while the rest stay near zero, as shown in Fig.~\ref{figure_observation}~(c). We trace this sparsity to the rate constraint: removing the bitrate term from the loss restores a balanced distribution, which confirms that the sparsity is rate-friendly and lowers the entropy for higher compression ratios~\cite{soap_lic}. However, common KD losses, such as mean squared error (MSE), frequency-domain distance, and Kullback-Leibler (KL) divergence, overlook this energy imbalance. A codec-oriented distillation loss is therefore needed.

Motivated by the above observations, we propose a two-stage ``NAS-then-KD'' neural video coding framework, named KD-NVC.
In the first stage, we introduce an acceleration-efficiency-based NAS~(AE-NAS) algorithm to determine the lightweight student architecture.
It explores module-wise Pareto frontiers to identify the acceleration capacity of each module and better allocate the acceleration budget across heterogeneous modules (for Obs.~I), and then determines the final student architecture through the estimated acceleration-efficiency metric, without practically training architecture-level candidates to save NAS time (for Obs.~II).
In the second stage, to address the sparse feature-energy distribution inherent in neural codecs, we design an energy-aware feature distillation (EFD) loss to improve the convergence efficiency and RD performance of the student codec (for Obs.~III).
Experimental results demonstrate that the proposed framework outperforms existing codec-oriented distillation methods under different acceleration settings. In particular, even when the model is streamlined to enable 69 frames per second (FPS) decoding at 1080p on an RTX 5060 GPU (over $2\times$ faster than the lightweight codec DCVC-RT~\cite{dcvc_rt}), our method still achieves compression performance comparable to VTM-LDB~\cite{vtm}, as shown in Fig.~\ref{figure_first}.

Our contributions are summarized as follows:

\begin{itemize}

    \item We propose KD-NVC, a two-stage distillation framework for decoder-side acceleration of neural video codecs, which includes efficient module-wise architecture selection and codec-oriented feature distillation.
    \item We introduce an acceleration-efficiency-based neural architecture search (AE-NAS) algorithm, which allocates the acceleration budget across heterogeneous modules and performs efficient architecture selection without training architecture-level candidates.
    \item We design an energy-aware feature distillation (EFD) loss that transfers rate-induced feature-energy patterns, and integrate it into a simple one-step training pipeline.
\end{itemize}

\section{Related Works}

\subsection{Neural Video Coding}

Most neural video coding methods~\cite{dvc, dcvc, dcvc_dc, dcvc_sdd, tmm2023deformable} typically adopt a two-step pipeline including inter-prediction and residual/context compression. 
They improve rate-distortion performance from several perspectives, such as inter-prediction~\cite{dcvc_sdd, dcvc_dc}, residual/context coding~\cite{dcvc, ecvc, dcvc_tcm}, entropy models~\cite{dcvc_hem, dcvc_dc, dcvc_rt}, and training strategies~\cite{dcvc_sdd, sheng2024prediction}. Many NVC methods, such as the ECVC series~\cite{ecvc} and the DCVC series~\cite{dcvc_dc, dcvc_fm, dcvc_rt}, demonstrate excellent RD performance. Generative adversarial networks and diffusion models are also widely used to improve perceptual quality at low bitrates~\cite{plvc, s2vc, i2vc}. To accelerate video codecs, Jia~\emph{et al.}~\cite{dcvc_rt} proposed DCVC-RT, which replaces explicit motion estimation, coding, and compensation with a lightweight implicit inter-prediction module. It achieves faster video coding and becomes the baseline for many NVC methods~\cite{s2vc, xiang2025}. Some other methods further reduce coding latency by applying model quantization~\cite{icnvc, quant_nvc, van2024mobilenvc, mobilecodec} and better parallel design~\cite{dcvc_uf}.

\subsection{Knowledge Distillation in Image and Video Coding}

KD transfers knowledge from a large teacher model to a compact student model, thereby enabling the lightweight student model to achieve better performance. Both logits-based~\cite{hinton2015distilling} and feature-based knowledge distillation~\cite{kd_fitnets} have been extensively adopted across various domains (e.g., computer vision, natural language processing). In low-level vision tasks such as image restoration and super-resolution, most existing methods rely on feature distillation~\cite{kd_sr_pdsrn, li2024knowledge}. In learned image compression (LIC), distillation has recently been used to improve coding speed. Fu~\emph{et al.}~\cite{fu2024tip, feds} proposed aligning both the distribution parameters in the entropy model~\cite{checkerboard_context} and the reconstructed image to improve the performance of the lightweight codec. Allemand~\emph{et al.}~\cite{prim} proposed simply aligning latent representations and reconstructed images to improve the RD performance of the student model; Chen~\emph{et al.}~\cite{iccv2025smodi} proposed removing explicit feature-distance-based distillation loss and using a stage-wise fine-tuning strategy to implicitly distill lightweight modules. Wang~\emph{et al.}~\cite{wang2026distilling} employed weight-sharing supernets for NAS of the image encoder and decoder, and conducted a progressive distillation strategy for training the complexity-variable image codec. Tatwawadi~\emph{et al.}~\cite{pico_apple} designed a lightweight image codec via two-step NAS, which partially trains $1,000$ candidates to choose a subset containing $20$ better candidates, and fully trains the subset for final architecture selection.

In neural video coding, many methods~\cite{free_vsc, smc++, sec_vcm, s2vc} employ feature distillation to help video codecs better preserve semantic information and suppress non-semantic content, while pretrained visual foundation models~\cite{swin_transformer, dinov2} serve as teacher models. Thus, the reconstructed frames are more suitable for downstream machine vision tasks. However, accelerating the video coding process by distillation remains underexplored.

\section{Methodology}

\begin{figure*}[!t]
    \centering
    \includegraphics[width=\linewidth]{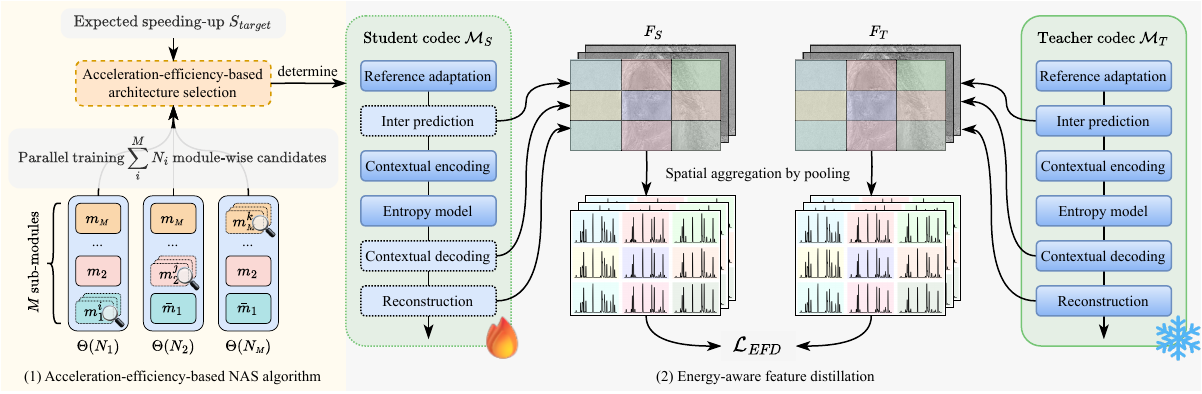}
    \caption{Pipeline of the proposed KD-NVC framework, which contains two stages. First, the architecture of the student codec $\mathcal{M}_S$ is determined by the proposed acceleration-efficiency-based NAS (AE-NAS) algorithm while avoiding training all architecture-level candidates. Then, energy-aware feature distillation (EFD) is performed from the perspective of channel-wise energy sparsity to improve the performance of the student $\mathcal{M}_S$.}
    \label{figure_main}
\end{figure*}

\subsection{Preliminaries and Framework Overview}

Formally, let $\{x_1,\dots,x_T\}$ denote a video sequence. A neural video codec encodes the current frame $x_t$ into a quantized latent $\hat{y}_t$ for entropy coding and reconstructs $\hat{x}_t$ using previously reconstructed frames $\hat{x}_{<t}$ as prior knowledge. 
Its training objective is the rate-distortion~(RD) loss $\mathcal{L}_{\textit{RD}}$:
\begin{equation}
    \mathcal{L}_{\textit{RD}} = \sum_{t=1}^T{[\mathcal{R}(\hat{y}_t|\hat{x}_{<t}) + \lambda \mathcal{D}(x_t, \hat{x}_t|\hat{x}_{<t})]}, 
\end{equation}

\noindent where $\mathcal{R}(\cdot)$ denotes the bitrate, $\mathcal{D}(\cdot, \cdot)$ denotes the distortion (e.g., MSE) between the original and reconstructed frames, and $\lambda$ controls the rate-distortion trade-off.

Building on the three observations in Section~\ref{section_observation}, we propose a two-stage ``NAS-then-KD'' framework, named KD-NVC, as shown in Fig.~\ref{figure_main}. 
In the first stage, instead of using handcrafted lightweight structures~\cite{fu2024tip, feds, iccv2025smodi, prim}, we formulate structure determination as a constrained search over lightweight architecture candidate space $\Omega$. We define the acceleration-efficiency $\eta$ as the speed-up obtained per unit of BD-rate degradation, and select the final architecture of student codec $\mathcal{M}_S$ that maximizes $\eta$ under a target speed-up ratio $S_{\textit{target}}$:

\begin{equation}
    \mathcal{M}_S = \mathop{\arg\max}_{\mathcal{M}^{\textit{arch}} \in \Omega}\,[\eta(\mathcal{M}^{\textit{arch}})] \quad \text{s.t.} \quad S(\mathcal{M}^{\textit{arch}}) \ge S_{\textit{target}},
    \label{eq:nas_objective}
\end{equation}

\noindent where $S(\mathcal{M}^{\textit{arch}})$ denotes the measured speed-up of the architecture-level candidate $\mathcal{M}^{\textit{arch}}$. Since exhaustively solving Eq.~\eqref{eq:nas_objective} is computationally prohibitive, we further divide it into two steps. We first explore the Pareto frontier of each module according to practical speed-up and RD degradation, and then restrict the final search to architecture-level candidates formed from these module-wise Pareto-frontier candidates. The estimated acceleration-efficiency $\hat{\eta}$ is used to rank architecture-level candidates and avoid exhaustively training them. This procedure is detailed in Section~\ref{section_nas}.

In the second stage, the selected student $\mathcal{M}_S$ is trained with both the original RD loss and the feature distillation term:

\begin{equation}
    \mathcal{L}_{\textit{total}} = \mathcal{L}_{\textit{RD}}(\mathcal{M}_S) + \beta \mathcal{L}_{\textit{EFD}}(F_T, F_S),
    \label{eq:total_loss}
\end{equation}

\noindent where $\mathcal{L}_{\textit{total}}$ denotes the total loss, $F_T$ and $F_S$ denote the intermediate features of the teacher codec $\mathcal{M}_T$ and student codec $\mathcal{M}_S$, respectively. $\beta$ controls distillation strength. $\mathcal{L}_{\textit{EFD}}$ denotes our proposed energy-aware feature distillation (EFD) loss, which aligns the student's features with the teacher's sparse feature-energy pattern for better RD performance. This procedure is detailed in Section~\ref{section_kdloss}.

\subsection{Acceleration-Efficiency-based Architecture Selection}
\label{section_nas}

\begin{figure}[t]
    \centering
    \includegraphics[width=\linewidth]{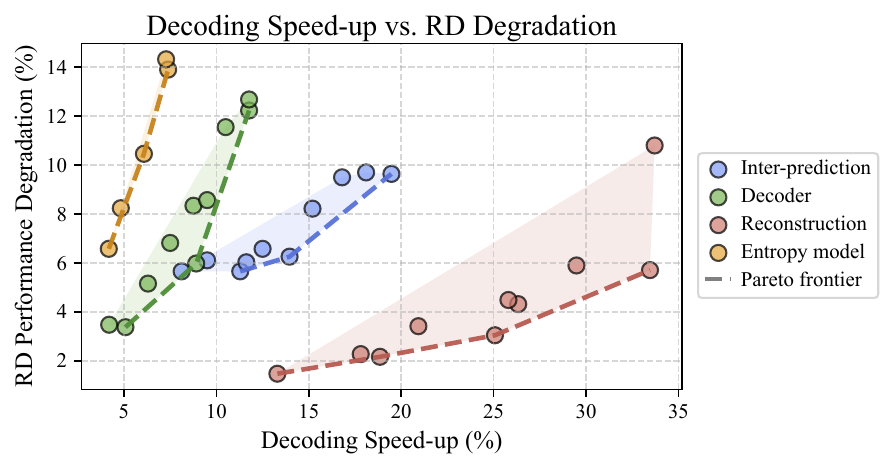}
    \caption{Decoding speed-up and rate-distortion performance degradation of different module structures. Different modules in the neural codec have distinct acceleration capacities.}
    \label{figure_acc_vs_bdrate}
\end{figure}

Different modules in neural video codecs contribute unevenly to the overall RD performance and computational latency, leading to different tolerances to model compression. As shown in Fig.~\ref{figure_acc_vs_bdrate}, some modules can be simplified with only minor RD penalty, while others are highly sensitive to structural simplification. Thus, uniform reduction across all modules may waste the acceleration budget on sensitive modules and result in a worse global performance-speed trade-off. To address this problem, we propose the acceleration-efficiency-based neural architecture search (AE-NAS) algorithm to determine the final per-module design of the student $\mathcal{M}_S$. Specifically, AE-NAS first obtains Pareto frontiers of each module through module-wise candidate training. Then, it ranks architecture-level candidates, which are formed by combining module-wise candidates, using the estimated acceleration-efficiency instead of practically training them.

Formally, let a neural video codec $\mathcal{M}$ be decomposed into a sequence of $M$ modules $\{m_1, m_2, \dots, m_M\}$. For each module $m_i$, we define a search space containing $N_i$ lightweight variants $\Omega_i=\{m_i^1, m_i^2, \dots, m_i^{N_i}\}$, where each variant changes channel, layer, or feed-forward network (FFN)~\cite{dcvc_rt} size. 

\begin{algorithm}[t]
\caption{Acceleration-Efficiency-based NAS Algorithm}
\label{alg:ae_nas}
\begin{algorithmic}[1]

\REQUIRE Teacher codec $\mathcal{M}_T=\{m_1,\dots,m_M\}$, $M$ module-wise variant spaces $\{\Omega_i,\dots,\Omega_M\}$, target speed-up $S_{\textit{target}}$
\ENSURE Selected student architecture $\mathcal{M}_S$

\STATE \textbf{\# Step 1: Module-wise search}
\FOR{each module $m_i, i=1\dots M$}
    \FOR{each lightweight variant $m_i^j \in \Omega_{i}$}
        \STATE $\mathcal{M}_{i}^{j}$ $\leftarrow$ Replace $m_i$ in $\mathcal{M}_T$ with $m_i^j$
        \STATE Fully optimize $\mathcal{M}_{i}^{j}$ with a short training strategy
        \STATE Measure RD drop $\Delta(m_i^j)$ using BD-rate
        \STATE Measure speed-up $S(m_i^j)$
    \ENDFOR
    \STATE Determine the Pareto-frontier candidate set $\mathbb{P}_{i}$ of $m_i$
\ENDFOR

\STATE \textbf{\# Step 2: Global architecture selection}
\STATE $\mathcal{M}_S \leftarrow \text{None}$
\STATE $\eta_{\text{max}} \leftarrow 0$
\STATE Construct the final search space $\Omega \leftarrow \mathbb{P}_{1} \times \cdots \times \mathbb{P}_{M}$
\FOR{each architecture candidate $\{\bar{m}_1,\dots,\bar{m}_M\} \in \Omega$}
    \STATE Construct $\mathcal{M}^{\textit{arch}} \leftarrow \{\bar{m}_1,\dots,\bar{m}_M\}$
    \STATE Estimate RD drop: $\hat{\Delta}(\mathcal{M}^{\textit{arch}}) \leftarrow \sum_{i=1}^{M}\Delta(\bar{m}_i)$
    \STATE Measure speed-up $S(\mathcal{M}^{\textit{arch}})$
    \STATE $\hat{\eta}(\mathcal{M}^{\textit{arch}}) \leftarrow S(\mathcal{M}^{\textit{arch}}) / \hat{\Delta}(\mathcal{M}^{\textit{arch}})$
    
    \IF{$S(\mathcal{M}^{\textit{arch}}) \geq S_{\textit{target}}$ \textbf{and} $\hat{\eta}(\mathcal{M}^{\textit{arch}}) > \eta_{\text{max}}$}
        \STATE $\eta_{\text{max}} \leftarrow \hat{\eta}(\mathcal{M}^{\textit{arch}})$
        \STATE $\mathcal{M}_S \leftarrow \mathcal{M}^{\textit{arch}}$
    \ENDIF
\ENDFOR

\RETURN $\mathcal{M}_S$

\end{algorithmic}
\end{algorithm}

In Step~1, each module $m_i$ is replaced one at a time by its lightweight variant $m_i^j$ ($1\le j \le N_i$), while other modules $m_{\neq i}$ remain identical to the teacher $\mathcal{M}_T$. In other words, each module-wise candidate $\mathcal{M}_i^j$ contains only one lightweight module. Then, we optimize all module-wise candidates in parallel to obtain the performance degradation $\Delta(m_i^j)$ induced by the $j$-th candidate of the $i$-th module. The speed-up $S(m_i^j)$ is measured on a single Tesla V100 with standard PyTorch operations, and the performance degradation is measured by the BD-rate~\cite{bd_rate} metric. In each module-wise candidate, only one lightweight module is trained from scratch while the other modules directly inherit weights from the teacher $\mathcal{M}_T$, so all candidates converge quickly with only $1/3$ iterations compared to the teacher codec's training strategy. After training all $\sum_{i=1}^{M} N_i$ module-wise candidates, the module-wise Pareto frontier sets $\{\mathbb{P}_{1}, \dots, \mathbb{P}_{M}\}$ are determined.

In Step~2, the final architecture-level search space $\Omega$ is formed as the Cartesian product (full combinations) of module-wise Pareto-frontier sets $\mathbb{P}_{1}\times\dots\times\mathbb{P}_{M}$. Since training all architecture-level candidates is impractical, AE-NAS ranks them using the estimated acceleration-efficiency metric $\hat{\eta}$. For each architecture-level candidate $\mathcal{M}^{\textit{arch}}$, the global speed-up $S(\mathcal{M}^{\textit{arch}})$ is directly measured, while the global BD-rate degradation $\hat{\Delta}(\mathcal{M}^{\textit{arch}})$ is estimated by accumulating per-module degradations $\Delta(\bar{m}_i)$. The above process is described as follows:

\begin{equation}
     \hat{\Delta}(\mathcal{M}^{\textit{arch}}) \approx \sum_{i=1}^{M}\Delta(\bar{m}_i),\quad \hat{\eta}(\mathcal{M}^{\textit{arch}}) = \frac{S(\mathcal{M}^{\textit{arch}})}{\hat{\Delta}(\mathcal{M}^{\textit{arch}})},
    \label{eq:acc_estimation}
\end{equation}

\noindent where $\bar{m}_i$ denotes the selected lightweight variant of the $i$-th module. Among architecture-level candidates satisfying $S(\mathcal{M}^{\textit{arch}})\ge S_{\textit{target}}$, AE-NAS selects the one with the largest $\hat{\eta}$ as the final student architecture $\mathcal{M}_S$. The above procedure is detailed in Algorithm~\ref{alg:ae_nas}.

\subsection{One-Step Energy-Aware Feature Distillation} 
\label{section_kdloss}

Neural video codecs exhibit energy sparsity in their intermediate features, where a small group of channels carries most information while many channels remain near zero~\cite{iccv2025smodi, he2022elic}. This sparsity helps reduce entropy and improves compression efficiency~\cite{soap_lic}. However, standard feature distillation losses (e.g., pixel-wise MSE, KL divergence) treat all locations uniformly and overlook this energy-sparsity characteristic. Thus, we propose the EFD loss to help the student learn the sparsity.

As shown in Fig.~\ref{figure_main}, EFD distills the channel-wise energy distribution from the teacher $\mathcal{M}_T$ to the student $\mathcal{M}_S$.
To relax the strict constraint of point-to-point spatial alignment and capture the sparse feature-energy signatures, we apply adaptive average pooling to obtain a compact and spatially-aggregated energy signature $E \in \mathbb{R}^{C \times k \times k}$, as follows:
\begin{equation}
    E = \mathcal{A}(F) = \textit{AdaptiveAvgPool}_{k\times k}(F),
    \label{eq5}
\end{equation}
where $F$ denotes the intermediate feature in teacher and student codecs.
The student is encouraged to learn the energy allocation pattern of the teacher through the EFD loss, which is formulated as:
\begin{equation}
    \mathcal{L}_{\textit{EFD}}(F_T,F_S) = \frac{1}{C \cdot k^2} \sum_{c=1}^{C} \sum_{j=1}^{k^2} \left\| E_T^{c,j} - E_S^{c,j} \right\|_2^2,
    \label{eq6}
\end{equation}
where $E_T^{c,j} = \mathcal{A}(F_T^{c,j})$ and $E_S^{c,j} = \mathcal{A}(F_S^{c,j})$ denote the spatially-aggregated energy signatures of the $c$-th channel and $j$-th patch in the teacher and student features, respectively. 

To avoid the training burden discussed in Observation II, we train the selected student codec $\mathcal{M}_S$ with a simple one-step distillation strategy. Unlike progressive training strategies~\cite{yu2024decoupling, wang2026distilling} or ``parallel-then-merge'' training strategies~\cite{iccv2025smodi}, one-step distillation optimizes all modules at the same time. We do not introduce any additional learnable adapters to align channel or spatial dimensions, since we find that they slow down the convergence speed in neural video codec training. The final objective of student $\mathcal{M}_S$ combines the RD loss~\cite{dvc, dcvc_rt} and the proposed EFD loss:

\begin{equation}
    \mathcal{L}_{\textit{total}}=\underbrace{R+\lambda D(x,\hat{x})}_{\mathcal{L}_{\textit{RD}}(\mathcal{M}_S)}+\beta\underbrace{\sum_{i=1}^M \mathcal{L}_{\textit{EFD}} (F_T^{m_i}, F_S^{\bar{m}_i})}_{\mathcal{L}_{\textit{EFD}}(\mathcal{M}_S, \mathcal{M}_T)},
    \label{eq7}
\end{equation}

\noindent where $\lambda$ denotes the Lagrange multiplier for rate-distortion balance, $\beta$ denotes the hyperparameter balancing distillation strength, $F_T^{m_i}$ and $F_S^{\bar{m}_i}$ denote the intermediate features in $i$-th module of teacher $\mathcal{M}_T$ and student $\mathcal{M}_S$, respectively.

\section{Experiments and Analysis}

\subsection{Experimental Settings}

\subsubsection{Datasets}

The Vimeo-90k~\cite{vimeo90k} dataset is used in early-stage training. It is a widely used dataset for NVC training and contains 64k training sequences with 7 frames. Following~\cite{dcvc_rt, xiang2025}, 8k raw Vimeo videos are cropped into a longer group of pictures (GOP) for fine-tuning. We evaluate our method on 6 datasets, including HEVC Classes B$\sim$E, UVG, and MCL-JCV~\cite{xiang2025, dcvc_rt, dcvc_dc}.

\subsubsection{Comparison Settings}

We implement existing LIC distillation methods for comparison. Specifically, Fu2024~\cite{fu2024tip} aligns both the probability distribution and the reconstruction-level output between the student and teacher codecs. PRIM~\cite{prim} employs one-step alignment of the latent representations and the reconstructed outputs. SMoDi~\cite{iccv2025smodi} removes feature distillation loss and conducts a ``parallel-then-merge'' pipeline. It directly optimizes each lightweight sub-module while other modules are kept frozen, and then jointly optimizes all lightweight modules. To ensure a fair comparison, all compared methods are trained based on the same training strategy. VTM-LDB-23.11~\cite{vtm} is used as the anchor to calculate the BD-rate~\cite{bd_rate}, which indicates bitrate change while achieving the same reconstruction quality. Lower BD-rate means more bitrate saving and better RD performance.

\subsubsection{Training Details of Teacher Codec $\mathcal{M}_T$}

\begin{table}[t]
    \setlength{\tabcolsep}{0.8pt}
    \renewcommand\arraystretch{1.2} 
    \centering
    \caption{Detailed training strategy of teacher and student codecs. Global batch size is set to 8.}
    \label{table_training_details}
    \setlength{\tabcolsep}{5pt}
    \begin{tabular}{cccccc}
        \toprule
        Stage & Learning rate & GOP & Patch & Epochs & Iters \\
        \midrule
        1 & $1e-4\rightarrow5e-6$ & $2\rightarrow4$ & $384\times256$ & 120 & 1.0M \\
        2 & $1e-4\rightarrow5e-6$ & $7$ & $384\times256$ & 25 & 200k \\
        3 & $1e-4\rightarrow5e-6$ & $15\rightarrow32$ & $384\times256$ & 230 & 230k \\
        4 & $1e-5\rightarrow5e-7$ & $64$ & $256\times256$ & 60 & 60k \\
        5 & $5e-7$ & $128$ & $256\times256$ & 30 & 30k \\
        \bottomrule
    \end{tabular}
\end{table}
 
Since the training script of DCVC-RT is not open-sourced, we reproduced it by following the original paper and incorporating advanced training strategies from recent NVC works. As shown in TABLE~\ref{table_training_details}, our training strategy includes 5 stages to gradually increase training GOP~\cite{dcvc_tcm, dcvc_sdd, sheng2024prediction}.
In the first stage, gradient descent is performed on each P-frame for pretraining and fast convergence, while the subsequent stages use the average gradient on all frames~\cite{dcvc_sdd, sheng2024prediction}. Vimeo-90k dataset is used in the first two stages, and raw Vimeo videos are used in the remaining three stages. To improve performance on long sequences, we applied hierarchical distortion weights $(0.9,1.2,0.9,0.5)$~\cite{dcvc_dc, dcvc_rt} and applied a random repetition strategy for the first P-frame~\cite{sheng2024prediction, avs_eem}. Furthermore, cyclical learning rates are adopted to avoid suboptimal local minima~\cite{xiang2025,circle_learning_rate}. Specifically, at the first 4 training stages, the learning rate progressively decreases, suddenly increases, and then progressively decreases.
Finally, to address GPU memory constraints during long sequence training, we employ the partial cascaded fine-tuning strategy~\cite{ecvc, sheng2024prediction} to reduce GPU memory usage. 
The Lagrange multiplier $\lambda$ for rate-distortion balance is set within $[1, 768]$~\cite{dcvc_rt}. The whole training process is performed on 8 Tesla V100 GPUs and takes 2 weeks.

\subsubsection{Training Details of Module-wise Candidates}

The training strategy of each module-wise candidate $\mathcal{M}_i^j$ is similar to that of $\mathcal{M}_T$, while epochs are reduced to $1/3$ of the above full training strategy. 

\begin{table}[t]
    \renewcommand\arraystretch{1.3}
    \centering
    \caption{Lightweight variants explored on the four modules. L\,/\,C\,/\,F denote retained ratios of layers / channels / FFN-expansion. Pareto-frontier variants of each module are in \colorbox{gray!20}{gray}.}
    \label{table_module_candidates}
    \setlength{\tabcolsep}{4pt}
    \begin{tabular}{@{} c m{0.74\linewidth} @{}}
        \toprule
        Module & Lightweight variants \\
        \midrule
        Entropy model &
            \colorbox{gray!20}{$\tfrac{1}{2}\text{F}$},\,
            $\tfrac{1}{2}\text{L}$,\,
            \colorbox{gray!20}{$\tfrac{1}{4}\text{F}$},\,
            \colorbox{gray!20}{$\tfrac{1}{2}\text{L}\tfrac{1}{2}\text{F}$},\,
            $\tfrac{1}{3}\text{L}$ \\
        \midrule
        Inter-prediction &
            $\tfrac{1}{2}\text{L}\,(2{:}1)$,\,
            \colorbox{gray!20}{$\tfrac{1}{2}\text{L}\,(1{:}2)$},\,
            \colorbox{gray!20}{$\tfrac{1}{2}\text{C}$},\,
            $\tfrac{1}{4}\text{F}$,\,
            $\tfrac{1}{2}\text{L}\tfrac{1}{2}\text{F}$,\,
            $\tfrac{1}{2}\text{F}$,\,
            \colorbox{gray!20}{$\tfrac{1}{2}\text{C}\tfrac{1}{4}\text{F}$},\,
            $\tfrac{1}{4}\text{C}$,\,
            $\tfrac{1}{2}\text{L}\tfrac{1}{2}\text{C}$,\,
            $\tfrac{1}{2}\text{L}\tfrac{1}{4}\text{F}$ \\
        \midrule
        Decoder &
            \colorbox{gray!20}{$\tfrac{2}{3}\text{L}$},\,
            \colorbox{gray!20}{$\tfrac{1}{2}\text{C}$},\,
            \colorbox{gray!20}{$\tfrac{2}{3}\text{L}\tfrac{1}{2}\text{F}$},\,
            $\tfrac{1}{2}\text{C}\tfrac{1}{2}\text{F}$,\,
            $\tfrac{1}{2}\text{F}$,\,
            $\tfrac{1}{4}\text{F}$,\,
            $\tfrac{1}{4}\text{C}$,\,
            $\tfrac{1}{3}\text{L}\tfrac{1}{2}\text{F}$,\,
            $\tfrac{2}{3}\text{L}\tfrac{1}{2}\text{C}$ \\
        \midrule
        Recon &
            \colorbox{gray!20}{$\tfrac{1}{2}\text{C}$},\,
            \colorbox{gray!20}{$\tfrac{1}{2}\text{L}\tfrac{1}{2}\text{C}$},\,
            \colorbox{gray!20}{$\tfrac{1}{2}\text{F}$},\,
            $\tfrac{1}{4}\text{F}$,\,
            $\tfrac{1}{2}\text{C}\tfrac{1}{2}\text{F}$,\,
            $\tfrac{1}{2}\text{L}\tfrac{1}{2}\text{F}$,\,
            $\tfrac{1}{2}\text{L}\tfrac{1}{4}\text{F}$,\,
            $\tfrac{1}{4}\text{C}$,\,
            $\tfrac{1}{2}\text{L}$,\,
            \colorbox{gray!20}{$\tfrac{1}{4}\text{L}$} \\
        \bottomrule
    \end{tabular}
\end{table}

\subsubsection{Training Details of Student Codec $\mathcal{M}_S$}

The pooling window size $k$ in Eq.~\eqref{eq5} and Eq.~\eqref{eq6} is set to $8$. The factor $\beta$ of the proposed $\mathcal{L}_{\textit{EFD}}$ in Eq.~\eqref{eq7} is set to $1.0$ in the first stage, and is then set to $0.0$ for faster convergence and better RD performance. Other details remain the same as those of the teacher codec $\mathcal{M}_T$.

\subsection{Student Architecture Selection}

\begin{table*}
    \centering
    \setlength{\tabcolsep}{3pt}
    \renewcommand\arraystretch{1.12}
    \caption{Rate-distortion performance (BD-rate \%) comparison of different feature distillation methods at $\approx$60\% and $\approx$100\% decoding speed-up in YUV420 color space. Speed-up is tested on 1080p video with a single NVIDIA Tesla V100-SXM2-32GB. \textbf{Bold} indicates the best result among distillation methods.}
    \label{tab:rd_performance}
    \resizebox{\textwidth}{!}{%
        \begin{tabular}{l|cccccc>{\columncolor{avgcolumnbg}}c|cccccc>{\columncolor{avgcolumnbg}}c}
            \toprule
            \multirow{2}{*}{Compared Methods} & \multicolumn{7}{c|}{IP32} & \multicolumn{7}{c}{IP-1} \\
            \cmidrule(lr){2-8} \cmidrule(lr){9-15}
             & UVG & HEVC-B & HEVC-C & HEVC-D & HEVC-E & MCL-JCV & Avg. & UVG & HEVC-B & HEVC-C & HEVC-D & HEVC-E & MCL-JCV & Avg. \\
            \midrule
            VTM-LDB-23.11~\cite{vtm} & 0.0 & 0.0 & 0.0 & 0.0 & 0.0 & 0.0 & 0.0 & 0.0 & 0.0 & 0.0 & 0.0 & 0.0 & 0.0 & 0.0 \\
            VVenC-LDB-1.15.0~\cite{vvenc}~$_{\textit{faster}}$ & 20.7 & 23.2 & 4.5 & 54.8 & 40.5 & 21.9 & 27.6 & 30.1 & 39.3 & 45.0 & 50.0 & 45.5 & 33.6 & 40.6 \\
            VVenC-LDB-1.15.0~\cite{vvenc}~$_{\textit{medium}}$ & -0.4 & -1.8 & -15.7 & 25.9 & 11.9 & -2.0 & 3.0 & 7.8 & 12.0 & 17.5 & 20.7 & 15.6 & 7.5 & 13.5 \\
            VVenC-LDB-1.15.0~\cite{vvenc}~$_{\textit{slower}}$ & -8.5 & -11.2 & -25.7 & 11.8 & 4.2 & -10.3 & -6.6 & -3.4 & -0.6 & 1.4 & 5.3 & 2.7 & -3.0 & 0.4 \\
            DCVC-FM (CVPR'24)~\cite{dcvc_fm} & -17.6 & -24.6 & -46.0 & -30.0 & -24.9 & -13.0 & -26.0 & -20.1 & -19.2 & -31.4 & -40.9 & -28.2 & -7.6 & -24.6 \\
            DCVC-RT (CVPR'25)~\cite{dcvc_rt} & -17.3 & -9.9 & -12.0 & -23.4 & -19.1 & -9.7 & -15.2 & -24.1 & -18.9 & -19.9 & -28.0 & -20.1 & -12.4 & -20.6 \\
            DCVC-RT (reproduced, $\mathcal{M}_T$) & -15.5 & -9.0 & -14.7 & -27.3 & -19.8 & -5.7 & -15.3 & -19.4 & -16.0 & -21.2 & -32.3 & -17.0 & -8.0 & -19.0 \\
            \midrule
            Distillation Methods & \multicolumn{14}{c}{\textit{$\approx$ 60\% decoding speed-up by AE-NAS algorithm}} \\
            \midrule
            Direct training & -3.0 & 2.4 & -3.0 & -18.2 & -15.5 & 5.1 & -5.4 & -4.9 & -0.2 & -5.5 & -20.3 & -0.1 & 7.0 & -4.0 \\
            Fu2024 (TIP'24)~\cite{fu2024tip} & -4.9 & 0.1 & -4.7 & -19.4 & -15.3 & 3.6 & -6.8 & -7.6 & -2.8 & -8.3 & -22.4 & 0.1 & 5.1 & -6.0 \\
            PRIM (ArXiv'25)~\cite{prim} & -5.2 & -0.3 & -5.4 & -19.9 & -15.7 & 3.4 & -7.2 & -8.0 & -4.0 & -9.2 & -22.6 & -1.1 & 4.3 & -6.8 \\
            SMoDi (ICCV'25)~\cite{iccv2025smodi} & -5.1 & 0.1 & -5.0 & -20.2 & -17.3 & 3.4 & -7.4 & -9.4 & -5.9 & -10.0 & -24.3 & -8.9 & 2.1 & -9.4 \\
            KD-NVC-S (Ours) & \textbf{-8.0} & \textbf{-1.7} & \textbf{-6.4} & \textbf{-20.7} & \textbf{-17.5} & \textbf{0.4} & \textbf{-9.0} & \textbf{-11.9} & \textbf{-7.7} & \textbf{-11.9} & \textbf{-24.8} & \textbf{-10.4} & \textbf{-0.1} & \textbf{-11.1} \\
            \midrule
            Distillation Methods & \multicolumn{14}{c}{\textit{$\approx$ 100\% decoding speed-up by AE-NAS algorithm}} \\
            \midrule
            Direct training & 6.2 & 11.3 & 6.9 & -9.6 & -10.6 & 13.7 & 3.0 & 7.9 & 12.5 & 8.0 & -8.9 & 16.3 & 18.4 & 9.0 \\
            Fu2024 (TIP'24)~\cite{fu2024tip} & 4.0 & 8.4 & 4.7 & -11.6 & -10.8 & 12.5 & 1.2 & 3.3 & 8.7 & 4.2 & -11.8 & 14.8 & 16.2 & 5.9 \\
            PRIM (ArXiv'25)~\cite{prim} & 3.4 & 7.6 & 4.1 & -11.9 & -10.8 & 11.9 & 0.7 & 2.1 & 7.3 & 2.8 & -13.3 & 12.6 & 15.1 & 4.4 \\
            SMoDi (ICCV'25)~\cite{iccv2025smodi} & 2.8 & 8.7 & 5.0 & -11.8 & -12.3 & 12.2 & 0.8 & 1.1 & 5.9 & 2.2 & -13.9 & 4.7 & 13.4 & 2.2 \\
            KD-NVC-T (Ours) & \textbf{0.1} & \textbf{6.3} & \textbf{2.0} & \textbf{-13.1} & \textbf{-13.1} & \textbf{9.1} & \textbf{-1.5} & \textbf{-1.3} & \textbf{4.1} & \textbf{0.2} & \textbf{-14.4} & \textbf{5.5} & \textbf{10.3} & \textbf{0.7} \\
            \bottomrule
        \end{tabular}%
    }
\end{table*}

\begin{figure}[t]
    \centering
    \includegraphics[width=\linewidth]{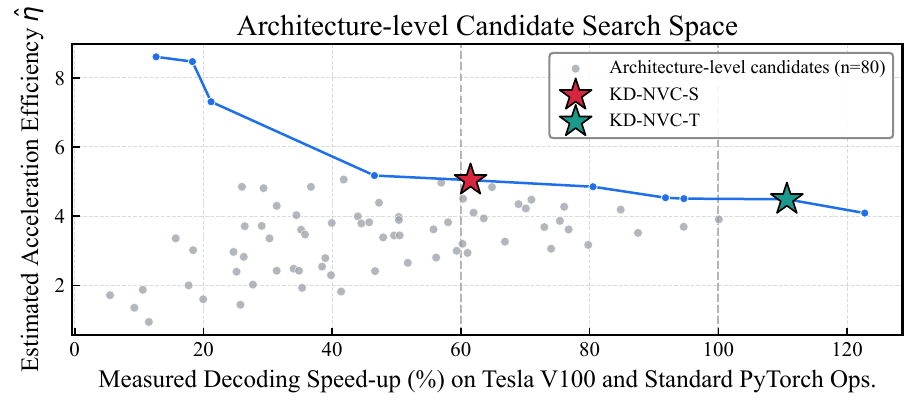}
    \caption{All architecture-level candidates in the final search space $\Omega$ of AE-NAS. The red and green star points indicate the selected student architectures under the $\approx 60\%$ and $\approx 100\%$ decoding speed-up targets, respectively.}
    \label{figure_estimated_ae}
\end{figure}

In the first step of AE-NAS, all $34$ module-wise candidates (in TABLE~\ref{table_module_candidates}) are trained with short training strategy. When channel reduction is applied, only the middle channels of stacked convolutions are reduced, while the input and output channels remain unchanged. This maintains the teacher-student dimension match, facilitating feature distillation.
After the first step, per-module Pareto-frontier candidates are determined, as shown in TABLE~\ref{table_module_candidates}, and used for the second step of global architecture selection. Since the entropy model is highly sensitive to simplification, it is excluded from the architecture-level candidate search space of AE-NAS. 
The search space $\Omega$ contains 80 candidates, and two architecture-level candidates are finally selected under two decoding speed-up targets, as shown in Fig.~\ref{figure_estimated_ae}. We denote the moderate $\approx 60\%$ speed-up student as KD-NVC-S, and the aggressive $\approx 100\%$ speed-up student as KD-NVC-T. Specifically, KD-NVC-S adopts inter-prediction~($\frac{1}{2}$L), decoder~($\frac{1}{3}$L), and reconstruction~($\frac{1}{2}$C). KD-NVC-T adopts inter-prediction~($\frac{1}{2}$L$\frac{1}{2}$F), decoder~($\frac{1}{3}$L$\frac{1}{2}$F), and reconstruction~($\frac{1}{2}$L$\frac{1}{2}$C). Both architectures are trained with the proposed one-step distillation paradigm and EFD loss.

\subsection{Rate-Distortion Performance}

We evaluate KD-NVC under Intra-Period (IP) $32$ and $-1$ by BD-rate, using VTM-LDB-23.11 as the anchor, as reported in TABLE~\ref{tab:rd_performance}. We reproduce DCVC-RT as the teacher codec. All compared distillation baselines use the same student architectures determined by AE-NAS, so the comparison purely focuses on the distillation paradigm. 
Under the moderate $\approx 60\%$ speed-up target, KD-NVC-S achieves the best average BD-rate in both IP settings: -9.0\% on IP32 and -11.1\% on IP-1. Compared with the SOTA distillation method SMoDi, it improves the average BD-rate by 1.6\% and 1.7\% on IP32 and IP-1, respectively.
Under the more aggressive $\approx 100\%$ speed-up target, KD-NVC-T outperforms VTM-LDB on IP32 (-1.5\%) and is comparable with VTM-LDB on IP-1 (0.7\%). 
The above experiments show the effectiveness of our one-step distillation with the proposed EFD loss. 

\subsection{Complexity Analysis}

\subsubsection{Theoretical Complexity}
The kMACs/pixel (computed by \texttt{fvcore}~\cite{fvcore}) at 1080p resolution and the total parameter count are shown in TABLE~\ref{tab:complexity_kmacs}. KD-NVC-S and -T reduce decoding complexity by $44\%$ and $57\%$, respectively. The parameter count drops from 20.7~M (DCVC-RT) to 16.0~M and 14.6~M for KD-NVC-S and -T, reducing by $22.5\%$ and $29.3\%$, respectively. 

\begin{table}[t]
    \renewcommand\arraystretch{1.2}
    \setlength{\tabcolsep}{3pt}
    \centering
    \caption{Total parameter count and computational complexity (kMACs/pixel) at 1080p resolution.}
    \label{tab:complexity_kmacs}
    \begin{tabular}{l|c|cc|cc}
        \toprule
        \multirow{2}{*}{Model} & \multirow{2}{*}{Params (M)} & \multicolumn{2}{c|}{kMACs/pixel} & \multicolumn{2}{c}{Reduction} \\
        & & Enc. & Dec. & Enc. & Dec. \\
        \midrule
        DCVC-FM~\cite{dcvc_fm} & 18.3 & 1138.1 & 866.7 & $+$699\% & $+$420\% \\
        DCVC-RT~\cite{dcvc_rt} & 20.7 & 142.4 & 166.8 & -- & -- \\
        KD-NVC-S & 16.0 & 109.2 & 93.7 & $-$23\% & $-$44\% \\
        KD-NVC-T & 14.6 & 93.7 & 71.7 & $-$34\% & $-$57\% \\
        \bottomrule
    \end{tabular}
\end{table}

\subsubsection{Practical Speed}
The FPS measured on desktop, server, and edge devices are shown in TABLE~\ref{tab:complexity_fps}. KD-NVC pushes DCVC-RT toward real-time or high-frame-rate decoding. It crosses the high-frame-rate (60 FPS) threshold at 1080p on both RTX 5060 and Tesla V100. At 2K resolution, DCVC-RT falls below real-time (30 FPS) on the RTX 5060 and Tesla V100, whereas KD-NVC achieves real-time decoding. On Apple MacBook Air M3, KD-NVC-S and -T also reach real-time 480p decoding. These results show that KD-NVC achieves practical FPS increase on diverse devices.

\begin{table*}[t]
    \renewcommand\arraystretch{1.2}
    \setlength{\tabcolsep}{3pt}
    \centering
    \caption{Encoding and decoding speed (FPS) on different hardware platforms and resolutions. Standard PyTorch operations are replaced with CUDA implementations on RTX 5060 and Tesla V100, and run on the MPS backend on Apple MacBook Air M3. \colorbox{green!10}{Light green} indicates achieving real-time 30 FPS; \colorbox{green!40}{dark green} indicates achieving high-frame-rate 60 FPS.}
    \label{tab:complexity_fps}
    \newcolumntype{C}{>{\centering\arraybackslash}p{1.35cm}}
    \newcolumntype{H}{>{\centering\arraybackslash}p{\dimexpr 2.7cm + 6pt\relax}|}
    \newcolumntype{G}{>{\centering\arraybackslash}p{\dimexpr 2.7cm + 6pt\relax}}
    \begin{tabular}{l|CC|CC|CC|CC|CC}
        \toprule
        \multirow{2}{*}{Model} & \multicolumn{2}{H}{RTX 5060 8G, 1080p} & \multicolumn{2}{H}{RTX 5060 8G, 2K} & \multicolumn{2}{H}{Tesla V100 32G, 1080p} & \multicolumn{2}{H}{Tesla V100 32G, 2K} & \multicolumn{2}{G}{Apple M3 16G, 480p} \\
        & Enc. & Dec. & Enc. & Dec. & Enc. & Dec. & Enc. & Dec. & Enc. & Dec. \\
        \midrule
        DCVC-FM~\cite{dcvc_fm} & OOM & OOM & OOM & OOM & 1.0 & 1.7 & 0.6 & 1.0 & 1.1 & 1.2 \\
        DCVC-RT~\cite{dcvc_rt} & \cellcolor{green!10}35.3 & \cellcolor{green!10}31.2 & 20.6 & 18.1 & \cellcolor{green!10}44.5 & \cellcolor{green!10}43.3 & 27.2 & 27.0 & 12.1 & 21.7 \\
        KD-NVC-S & \cellcolor{green!10}47.2 & \cellcolor{green!10}53.2 & 27.2 & \cellcolor{green!10}30.2 & \cellcolor{green!10}51.7 & \cellcolor{green!40}61.2 & \cellcolor{green!10}32.8 & \cellcolor{green!10}38.7 & 15.8 & \cellcolor{green!10}30.3 \\
        KD-NVC-T & \cellcolor{green!10}54.8 & \cellcolor{green!40}69.2 & \cellcolor{green!10}30.9 & \cellcolor{green!10}39.2 & \cellcolor{green!10}55.6 & \cellcolor{green!40}69.8 & \cellcolor{green!10}34.4 & \cellcolor{green!10}44.6 & 18.6 & \cellcolor{green!40}36.0 \\
        \bottomrule
    \end{tabular}
\end{table*}

\subsubsection{Theoretical and Practical Complexity}

\begin{figure}[!t]
    \centering
    \includegraphics[width=\linewidth]{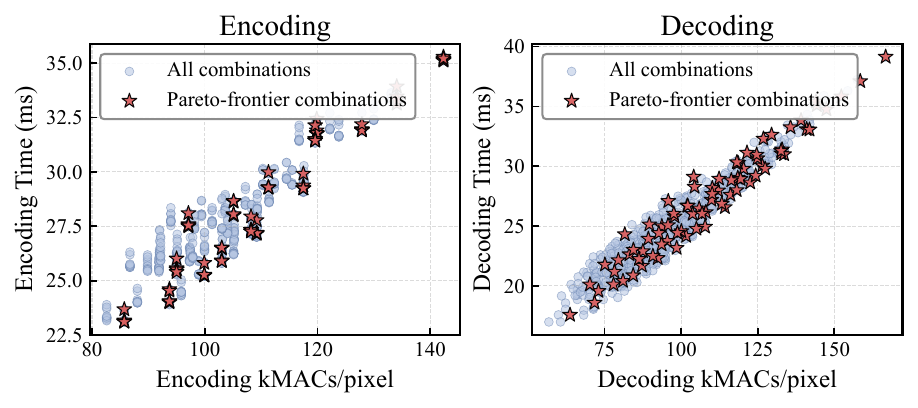}
    \caption{Theoretical complexity (kMACs/pixel) versus practical latency (ms) on 1080p for all possible lightweight architectures. Two subplots denote encoding and decoding processes, respectively. Latency is measured on a single NVIDIA Tesla V100-SXM2-32GB GPU.}
    \label{figure_kmacs_vs_time}
\end{figure}

We visualize the relationship between kMACs/pixel and practical latency for all $704$ architecture-level candidates and the $80$ Pareto-based architecture-level candidates searched by AE-NAS. Fig.~\ref{figure_kmacs_vs_time} shows that our searched Pareto frontier subset effectively offers sufficient architecture diversity and covers a broad coding latency range.

\subsection{Ablation Study}

\subsubsection{Ablation on AE-NAS Algorithm}

\begin{figure}[!t]
    \centering
    \includegraphics[width=\linewidth]{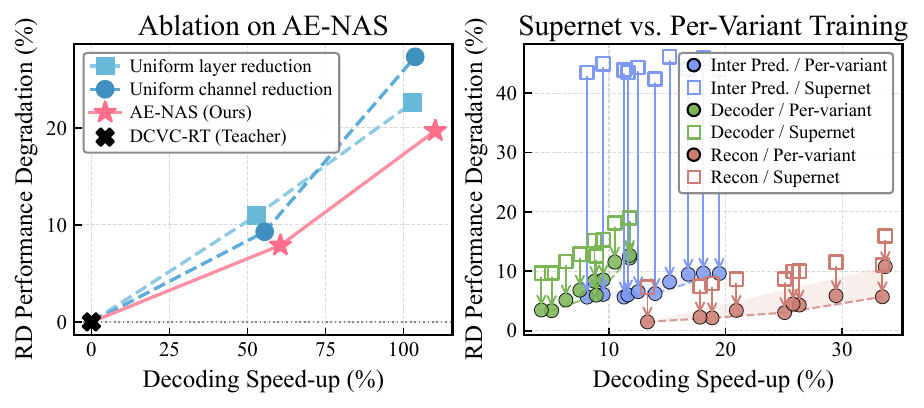}
    \caption{Left: Comparison between uniform architecture reduction and the proposed AE-NAS algorithm. Right: Comparison between supernet-based training and per-variant training. We observe that the supernet converges poorly, which underestimates the capacity of the inter-prediction module.}
    \label{figure_ablation_supernet}
\end{figure}


To validate the effectiveness of AE-NAS, the uniform model layer and channel reduction are used as comparison methods. For a fair comparison, reduction is applied to the inter-prediction, decoder, and reconstruction modules. As shown in Fig.~\ref{figure_ablation_supernet}, experimental results indicate that the per-module design of the AE-NAS algorithm achieves better RD performance.

The weight-sharing supernet is an efficient approach to rank lightweight variants in image compression~\cite{wang2026distilling}. It integrates multiple module variants into a single network to save training time. However, the ranking among different NVC modules cannot be accurately reflected by the supernet-based training. Our experiments find that the supernet increases the training difficulty and results in poor convergence of the inter-prediction module, which leads to an erroneous conclusion that the inter-prediction module has poor acceleration potential, as shown in Fig.~\ref{figure_ablation_supernet}. Therefore, we abandon the supernet-based training, and adopt individual training for all module-wise candidates to obtain the more accurate per-variant rate-distortion-speed performance.


\subsubsection{Ablation on Proposed Energy-aware KD Loss}

\begin{figure}[t]
    \centering
    \includegraphics[width=\linewidth]{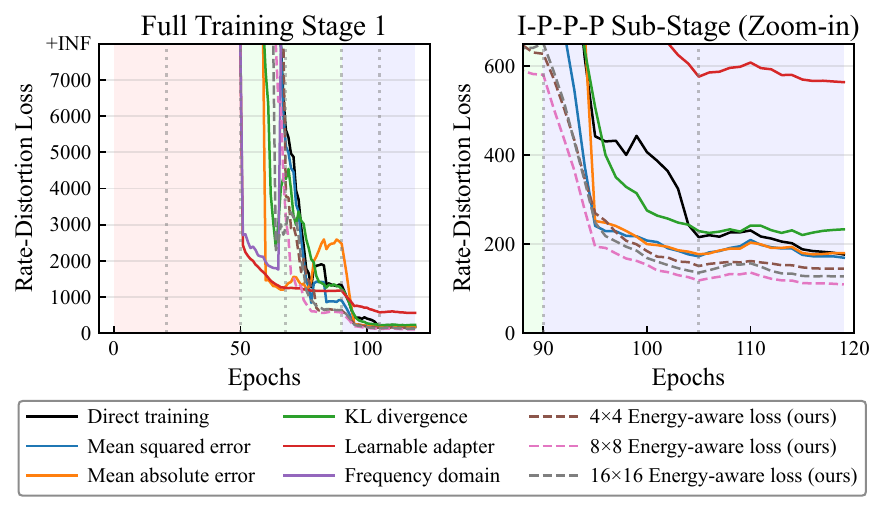}
    \caption{Rate-distortion performance degradation under different distillation losses in the first stage, which consists of about $1,000,000$ iterations. The anchor is the fully trained DCVC-RT. \textcolor{myred}{Red}, \textcolor{mygreen}{green}, and \textcolor{myblue}{blue} regions indicate the I-P, I-P-P, and I-P-P-P training sub-stages, respectively. Vertical gray dashed lines denote start epochs of cyclical learning rates, where learning rates suddenly increase to avoid local minima. }
    \label{figure_loss_comparison}
\end{figure}

\begin{table}[t]
    \centering
    \setlength{\tabcolsep}{2.5pt}
    \renewcommand\arraystretch{1.2} 
    \caption{Comparison of training cost of different feature-distillation loss functions on a single $384\times256$ P-frame patch.}
    \label{tab:loss_compare}
        \begin{tabular}{lcc}
        \toprule
        Distillation loss & Iteration time (ms) & GPU memory (MiB) \\
        \midrule
        Direct training & 69.3 & 1,721 \\
        MSE~\cite{prim, kd_fitnets, fu2024tip} & 86.7 & 1,816 \\
        MAE~\cite{liu2024sar_kd} & 87.2 & 1,969 \\
        KL divergence~\cite{unicompress, yu2024decoupling} & 86.4 & 1,822 \\
        Learnable adapter~\cite{liu2023simple} & 88.9 & 1,839 \\
        Frequency domain~\cite{pham2024frequency_fam_kd} & 482.1 & 22,093 \\
        Energy-aware loss (Ours) & 85.4 & 1,816 \\
        \bottomrule
    \end{tabular}
\end{table}

To validate the proposed EFD loss, we compare it with common feature-distillation metrics under the same KD-NVC-S architecture and the same training strategy. The baselines include direct training, MSE~\cite{prim, kd_fitnets, fu2024tip}, MAE~\cite{liu2024sar_kd}, KL divergence~\cite{unicompress, yu2024decoupling}, learnable adapters~\cite{liu2023simple}, and frequency-domain adapters~\cite{pham2024frequency_fam_kd}. Their time and memory costs are detailed in TABLE~\ref{tab:loss_compare}. Because full NVC training is expensive, Fig.~\ref{figure_loss_comparison} shows the first training stage with about 1,000,000 iterations. We observe three findings. First, the adapter-based and KL-based losses even hurt convergence and fall behind direct training, which is consistent with our analysis since both metrics completely ignore the inter-channel energy disparity inherent to NVC features. Second, MAE, MSE, and frequency-domain alignment offer no convergence benefit over direct training, indicating that uniformly matching every point, every channel, or every frequency component does not transfer useful knowledge from teacher to student. Third, the student with proposed EFD converges faster and is significantly better than all the other losses throughout the training stage, confirming that aligning channel-wise energy signatures provides a more effective optimization signal for distilling the video codec.

\subsubsection{Ablation on One-Step Distillation}

To validate the one-step distillation paradigm, we compare it with direct training (no distillation), two progressive distillation paradigms (head-to-tail and tail-to-head)~\cite{wang2018progressive, wang2026distilling}, and a parallel distillation paradigm~\cite{iccv2025smodi}, all using the same KD-NVC-S architecture. Taking the one-step paradigm as the anchor ($0.0\%$ bitrate change, $1\times$ training GPU days cost), direct training increases the bitrate by $4.5\%$ at about $0.8\times$ training cost; head-to-tail and tail-to-head progressive distillation slightly increase the bitrate by $0.4\%$ and $0.3\%$ at about $3\times$ training cost; and parallel distillation increases the bitrate by $1.9\%$ with about $1.6\times$ training cost. The simple one-step paradigm therefore yields the best bitrate among all paradigms while keeping training cost close to or lower than the alternatives.

\subsection{Visualization}

\begin{figure*}[t]
    \centering
    \setlength{\lineskip}{0pt}
    \setlength{\parskip}{0pt}
    \includegraphics[width=\linewidth]{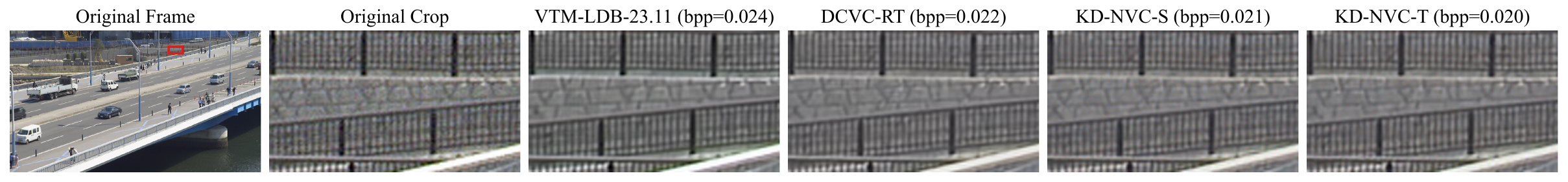}\\[-2pt]
    \includegraphics[width=\linewidth]{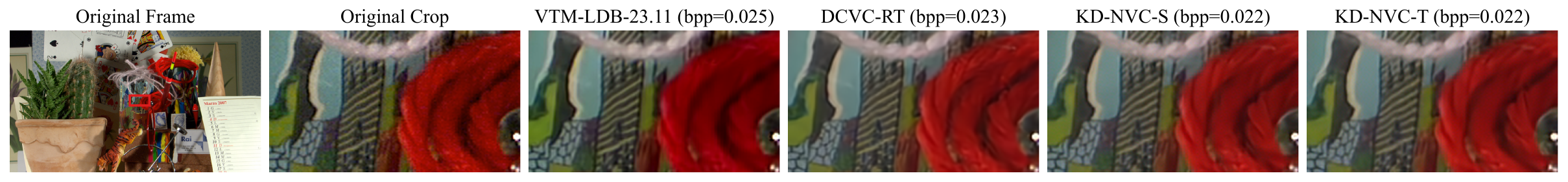}\\[-2pt]
    \includegraphics[width=\linewidth]{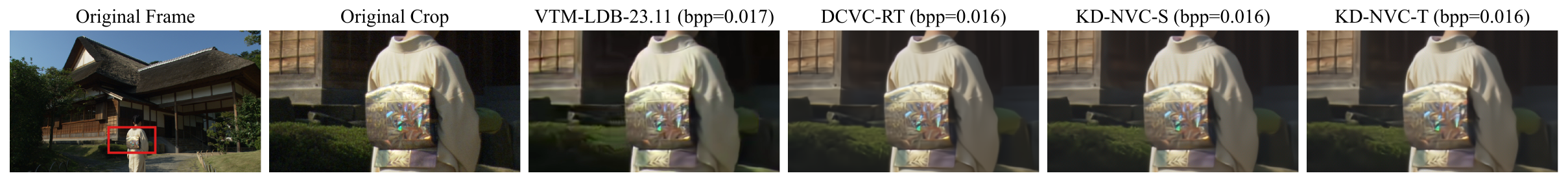}
    \caption{Visualization on HEVC test sequences of the original frame, VTM-LDB-23.11, the teacher codec DCVC-RT, and the proposed KD-NVC-S/T. KD-NVC achieves comparable visual quality at a very low bitrate (bpp $\approx 0.02$).}
    \label{figure_subjective}
\end{figure*}

Subjective comparisons on HEVC test sequences are shown in Fig.~\ref{figure_subjective}. For each sequence, we show the original frame and the reconstructed results by the VTM-LDB-23.11, the teacher codec DCVC-RT, and the proposed KD-NVC-S/T, with the corresponding bpp. Despite operating at a very low bitrate (bpp $\approx 0.02$), KD-NVC-S and -T preserve fine textures, structures, and other visual details. This confirms that KD-NVC retains the perceptual fidelity of the teacher while delivering substantial decoding speed-ups.

\section{Conclusion}

In this paper, we proposed KD-NVC, a two-stage ``NAS-then-KD'' framework to accelerate the decoding process of the neural video codec. KD-NVC addresses three challenges: heterogeneous module sensitivity, prohibitive NAS cost, and codec-specific feature-energy sparsity. The proposed AE-NAS algorithm efficiently selects lightweight student architectures by maximizing estimated acceleration-efficiency under speed-up constraints, while the proposed EFD loss aligns channel-wise energy patterns to transfer critical knowledge from teacher codec to student codec. Experiments show KD-NVC consistently improves the FPS/BD-rate trade-off at both $\approx60\%$ and $\approx100\%$ decoding speed-up settings. We hope this work can provide insights for future research in lightweight neural video coding. 

In the future, we will explore integer-based inference, quantization-aware training, parallel pipeline design, and asymmetric ``one fast decoder and multiple encoders'' design for practical application.

\bibliographystyle{IEEEtran}
\bibliography{references}

\end{document}